\documentstyle[12pt]{article}

\catcode`@=11

\def\@citex[#1]#2{\if@filesw\immediate\write\@auxout{\string\citation{#2}}\fi
  \def\@citea{}\@cite{\@for\@citeb:=#2\do
    {\@citea\def\@citea{,\penalty\@m}\@ifundefined
      {b@\@citeb}{{\bf ?}\@warning
       {Citation `\@citeb' on page \thepage \space undefined}}%
\hbox{\csname b@\@citeb\endcsname}}}{#1}}

\def\citer{\@ifnextchar [{\@tempswatrue\@citexr}{\@tempswafalse\@citexr[]}}

\def\@citexr[#1]#2{\if@filesw\immediate\write\@auxout{\string\citation{#2}}\fi
  \def\@citea{}\@cite{\@for\@citeb:=#2\do
    {\@citea\def\@citea{--\penalty\@m}\@ifundefined
       {b@\@citeb}{{\bf ?}\@warning
       {Citation `\@citeb' on page \thepage \space undefined}}%
\hbox{\csname b@\@citeb\endcsname}}}{#1}}
\catcode`@=12

\def\u{\upsilon}

\def\bo{{\raise.15ex\hbox{\large$\Box$}}}               
\def\face{{\raise.2ex\hbox{$\displaystyle \bigodot$}\mskip-2.2mu \llap {$\ddot
        \smile$}}}                                      

\def\leftrightarrowfill{$\mathsurround=0pt \mathord\leftarrow \mkern-6mu
        \cleaders\hbox{$\mkern-2mu \mathord- \mkern-2mu$}\hfill
        \mkern-6mu \mathord\rightarrow$}       
\def\dvec#1{\vbox{\ialign{##\crcr
        \leftrightarrowfill\crcr\noalign{\kern-1pt\nointerlineskip}
        $\hfil\displaystyle{#1}\hfil$\crcr}}}           

\def\beq{\begin{equation}}
\def\eeq{\end{equation}}

\def\beqx{\begin{displaymath}}
\def\eeqx{\end{displaymath}}

\def\beql{\begin{eqnarray}}
\def\eeql{\end{eqnarray}}

\begin{document}	

\begin{flushright}
NIKHEF/99-027\\
November 1999
\end{flushright}

\vspace{15mm}
\begin{center}
{\Large\bf\sc Open Descendants of Non-diagonal Invariants}
\end{center}
\vspace{2cm}
\begin{center}
{\large L.R Huiszoon, A.N. Schellekens, N.Sousa}\\
\vspace{15mm}
{\it NIKHEF Theory Group\\
P.O. Box 41882, 1009 DB Amsterdam, The Netherlands} \\
\end{center}

\vspace{2cm}

\begin{abstract}

The open descendants of simple current automorphism invariants are
constructed. We consider the case where the order of the current is two or
odd. We prove that our solutions satisfy the completeness conditions,
positivity and integrality of the open and closed sectors and the Klein bottle constraint (apart from an interesting exception). In order to do this, we derive some new relations between the tensor $Y$ 
and the fixed point conformal field theory. 
Some non-standard Klein
bottle projections are considered as well.

\end{abstract}

\thispagestyle{empty}

\newpage \setcounter{page}{2}

\section{Introduction}

After a ten year period of neglect, open strings have received more interest
recently for a variety of reasons: their r\^ole in the duality picture,
the discovery of D-branes and the appearance of non-commutative geometry,
and recent developments in phenomenological string theory, such as  brane
world scenarios and large extra dimensions.
It is therefore worthwhile to explore open string vacua in more detail.

A first step in the classification of open string theories is a
classification of closed string theories. This amounts to classifying all
possible modular invariants. There exists a systematic way of building
modular invariants using simple currents~\cite{simple}. This turns out to be very powerful:
apart from the charge conjugation invariant and a few sporadic
exceptions, all modular invariants can be built with the use of simple currents.This suggests the possibility that also in the
open string case simple currents should play a
crucial r\^ole in a systematic approach. Furthermore simple current
properties are not specific for one particular type of conformal
field theories (such as WZW models), but they are generic. It is precisely a
generic CFT description of open strings that we are after.

A modular invariant can be of `automorphism' type or `extension' type or
products thereof. The first ones are permutations of the primary fields that leave the fusion rules invariant. Among them are the diagonal and charge
conjugation invariant. Theories described by an extension invariant always
contain extra currents that extend the chiral algebra. The modular invariant of this extension is then again of automorphism type.
This extension procedure does not raise any new problem in open string theories.
Therefore we only need to consider pure automorphism invariants.

A second ingredient in the construction of open descendants is the
classification of boundary conditions for a given bulk theory. A bulk theory is specified by a chiral algebra and a particular modular invariant that
describes the pairing between left- and rightmoving representations of this algebra. The boundary conditions are encoded in the `boundary coefficients' $B_{b\alpha}$. These have the following intuitive interpretation. When a
field $b$ approaches a boundary $\alpha$, it gets reflected to its charge
conjugate $b^c$ with a strength proportional to $B_{b\alpha}$. Throughout
this letter, we will assume that the boundaries leave the bulk symmetry
invariant.

In the case that all fields couple to their charge conjugate, Cardy~\cite{cardy} conjectured 
the boundary coefficients for a generic conformal field
theory. We will refer to this as the C-diagonal case. 
In general, the boundary
coefficients are constrained by ``sewing
constraints''~\cite{sewing}~\cite{completeness}. Unfortunately, these
constraints require knowledge of OPE-coefficients and duality
(fusing and braiding)
matrices, which are only known in a limited number of cases, such as $SU(2)$
 WZW models. The authors of~\cite{completeness}~\cite{nondiagonal} found the boundary coefficients for the ``$D_{\rm odd}$" automorphism invariants of 
$SU(2)$ WZW models
 by solving these sewing constraints. Furthermore, they postulated very restrictive conditions for the annulus
coefficients (open string state multiplicities) known as the `completeness
conditions'. These conditions have the advantage that they do not involve
duality matrices, so that they are applicable in all cases.
Fuchs and Schweigert~\cite{class} generalized this
to the boundary coefficients for an  arbitrary order $2$, half-integer spin simple current
invariant. They constructed the corresponding annulus coefficients and showed that these were integer and satisfied the completeness conditions.

A third ingredient in the construction of open descendants are the Klein
bottle and M\"obius strip amplitudes. In short, the Klein bottle projects the closed string spectrum, described by the torus, to an unoriented one and the M\"obius strip does the same with the open spectrum, which is described by
the annulus.
The description of Klein Bottle and M\"obius strip requires the introduction
of another set of quantities,
the `crosscap coefficients' $\Gamma_b$, which together with
the boundary coefficients $B_{b\alpha}$ form the set of data needed for
a complete description of open string spectra.
The introduction of crosscaps to the string worldsheet leads to extra sewing constraints, the so-called `crosscap constraints' \cite{crosscap}.
These again require knowledge of the -- in general unknown -- duality matrices.
Since general formulas for duality matrices are not likely to be available
soon, any attempt to arrive at a generic CFT description will have to be
less ambitious. Instead of trying to solve all consistency
 conditions we will
focus here on a non-trivial but accessible subset. This is somewhat
analogous to the situation with closed string construction.
Although in principle one would have to check duality of the four-point
function on the sphere, in practice it is usually sufficient, and much
simpler, to check modular invariance on the torus.
The conditions we will consider in this paper are the aforementioned
completeness conditions, positivity and integrality of the open
and closed string partition functions, and a `Klein bottle constraint',
formulated in~\cite{descendants}, as an alternative
to the crosscap constraint. We require these constraints to be satisfied
for any CFT, thus focusing on those solutions that are most likely to
survive the full set of sewing constraints. Since our solutions
are simple current related there may in fact be a chance that simple
current relations among the duality matrices are sufficient to prove
that all sewing conditions are satisfied, if they are satisfied in the
C-diagonal case.

Until now, only in the C-diagonal case there exists a general prescription
to construct open descendants. This is based on the conjecture of Cardy,
extended with a conjecture for the crosscap. It
was generalized to a class of simple current
related crosscap coefficients, and it was shown that all consistency
conditions mentioned in the previous paragraph are satisfied.
In this letter we will provide the
prescription for the construction of open descendants in case of pure
automorphism modular invariants built from odd order and order $2$ simple
currents.

The rest of this paper is organized as follows: In section~\ref{sec-open}
we review the method of open descendants. In section~\ref{sec-odd} we
present solutions of the consistency conditions for automorphism invariants built from odd order simple currents. This case is relatively simple because of the absence of short orbits. Non-standard Klein bottle projections are
discussed in a subsection. Section~\ref{sec-two} is devoted to the
construction of open descendants of order $2$ simple current automorphism
invariants. Due to the possibility of fixed points, this case is harder to
deal with and we find that the correct crosscap coefficient requires a subtle sign choice. In some rare cases, the naive version of the Klein bottle
constraint is violated and we comment on a possible solution. Some
essential relations involving various types of 
generalized fusion coefficients
are derived in appendix $A$.

\section{Open descendants} \label{sec-open}

In this section we review the method of open descendants, based on the
original work of~\cite{sagnotti}~\cite{cardy}~\cite{descendants}. The closed sector is described by a modular invariant torus partition function
\begin{equation} \label{eq:torus}
	T = \sum_{ij} Z_{ij}\chi_{i} \bar{\chi}_{j}  \;\; ,
\end{equation}
where the index $i$ labels the (primary) fields and the matrix $Z_{ij}$ is
the modular invariant that commutes with the modular transformations $S$ and $T$. We will only be interested in the case where $Z_{ij}$ is a pure
automorphism, so that it has entries that are $0$ or $1$. A bar on the
characters $\chi_i$ denotes ``anti-holomorphic". We will introduce a name for the fields that couple to their charge conjugate in the torus: {\em
transverse channel fields}, or transverse fields for short, and denote them by $a,b$. This name is justified since only these fields can survive near a crosscap or boundary, and therefore only these fields can propagate in the
transverse channels.

The closed oriented theory is projected to an unoriented theory by the Klein bottle partition function $K$. This partition function,
or direct channel amplitude, is related by a channel transformation $S$ to
the transverse channel $\tilde {K}$:
\begin{equation}\label{eq:transK}
	K = \sum_i K_i \chi_i \;\; \overrightarrow{\;\;{}_S\;\;\;\;} \;\;
\tilde{K} = \sum_i \Gamma_i^2 \chi_i \;\;\;\;\; {\rm where} \;\;\; K_i =
\sum_b S_{bi} \Gamma_b \Gamma_b \;\; .
\end{equation}
The Klein bottle has to satisfy two consistency conditions. The first
condition guarantees that the closed string sector $(T+K)/2$ has positive,
integral state multiplicities:
\begin{equation} \label{eq:posintK}
	K_i = \epsilon_i Z_{ii}\;\; ,
\end{equation}
where  $\epsilon_i=\pm1$ for $Z_{ii} = 1$ and $\epsilon_i = 0$ for $Z_{ii} = 0$. These signs have to satisfy an additional constraint, namely
\begin{equation} \label{eq:epsilon}
	\epsilon_i \epsilon_j \epsilon_k N_{ij}^{~~k} \geq 0 \;\; ,
\end{equation}
where $N_{ij}^{~~k}$ are the fusion coefficients given by Verlinde's
formula~\cite{verlinde}. This condition, which we will refer to as the Klein bottle constraint, guarantees that states that are projected out by the Klein bottle cannot re-appear as intermediate states in tree diagrams.

The open sector is described by the direct annulus channel $A$ and direct
M\"obius channel $M$. They are related to the transverse channels $\tilde{A}$ and $\tilde{M}$ by a channel transformation, which is $S$ in case of the
annulus and $P=\sqrt{T}ST^2S\sqrt{T}$ for the M\"obius strip:
\begin{eqnarray}
		A_{\alpha\beta} =  \sum_{i} A_{i\alpha\beta} \chi_{i}  \;
&\overrightarrow{\;\;{}_S\;\;\;\;}& \;  	 \tilde{A}_{\alpha\beta}  =  \sum_b B_{b\alpha} B_{b\beta} \chi_b  \;\;\; \\ &&{\rm where} \;\;\;  A_{i\alpha\beta}  =  \sum_b S_{bi} B_{b\alpha} B_{b\beta}\label{eq:transA} \;\; ,\nonumber\\
		M_{\alpha} =  \pm  \sum_{i} M_{i\alpha} \hat{\chi}_i  \;
&\overrightarrow{\;\;{}_P\;\;\;\;}& \; 
 \tilde{M}_{\alpha} = \pm \sum_{b} \Gamma_b B_b \hat{\chi}_b \;\;\; \\&&{\rm
where} \;\;\; M_{i\alpha}  =  \sum_b P_{bi} B_{b\alpha} \Gamma_b \nonumber \;\;
,\label{eq:transM}
\end{eqnarray}
where the hatted characters are defined by $\hat{\chi_i} =(\sqrt{T_i})^{-1} \chi_i$. The indices $\alpha,\beta$ label the boundary conditions. To ensures that the open sector, given by $(A+M)/2$ has non-negative, integer state
degeneracies, the annulus and M\"obius strip coefficients have to satisfy
\begin{equation} \label{eq:posintM}
	|M_{i\alpha}| \leq A_{i\alpha\alpha} \;\; {\rm and} \;\; M_{i\alpha} = A_{i\alpha\alpha} \;\; {\rm mod} \, 2 \;\; .
\end{equation}
We will refer to this condition as the positivity and integrality condition. Furthermore, the annulus coefficients have to satisfy the completeness
conditions of~\cite{completeness}. This is automatically the case if the
reflection coefficients $R_{b\alpha}=B_{b\alpha}\sqrt{S_{b0}}$ satisfy~\cite{zuber}
\begin{equation} \label{eq:complcond}
	\sum_{b} R_{b\alpha}R_{b\beta}^*=\delta_{\alpha\beta}\ ; \ \ \
 	\sum_{\alpha} R_{b\alpha}R_{c\alpha}^*=\delta_{bc}\ . \end{equation}
These conditions will not be considered anymore, since they do not involve
the crosscap coefficients. That they are satisfied follows from the
results of \cite{class}, from which we obtain our boundary coefficients.

To summarize: In the construction of open descendants, for a given modular invariant and consistent Klein bottle projection, we have to find the
correct annulus and M\"obius coefficients. This amounts to finding a set of boundary labels $\alpha$ and
 coefficients
$B_{b\alpha}$ and $\Gamma_b$ such that
 equations~(\ref{eq:posintK}),~(\ref{eq:epsilon}),~(\ref{eq:posintM})
and~(\ref{eq:complcond})
are satisfied. Let us present the only model-independent
solution that is presently known, in its most general form.
It holds
for charge conjugation invariants,
the so-called C-diagonal~\cite{cardy} case.
 The relevant coefficients are in this case~\cite{klein}
\begin{eqnarray}
	&&B_{b\alpha} = \frac{S_{b\alpha}}{\sqrt{S_{bL}}} \label{CaRdY}
\;\;\;\; , \;\;\;
	\Gamma_{b} = \frac{P_{bL}}{\sqrt{S_{bL}}} \;\;\: , \;\;\; \\
 	&&K_i  = Y_{iLL^c} \;\;\; , \;\;\;
	A_{i\alpha\beta}  = N_{L^ci,\alpha\beta} \;\;\; , \;\;\;
	M_{i\alpha}  =  Y_{L^c\alpha,Li}  \;\;\; ,
\end{eqnarray}
where $L$ is a simple current and
the tensor $Y$ is defined as
\begin{equation} \label{eq:verlindeY}
	Y_{ij}^{\;\;\;k} = \sum_{m} \frac{S_{mi} P_{mj} P_{m}^k}{S_{m0}} \;\; .
\end{equation}
A crucial relation, derived in \cite{klein} using results
from ~\cite{bantay}, is
\begin{equation} \label{eq:bound}
	|Y_{i0}^{\;\;k} | \leq N^k_{\;ii} \;\;\;\; {\rm and} \;\;\;
Y_{i0}^{\;\;k} = N^k_{\;ii} \  {\rm mod} \, 2 \;\; .
\end{equation}
In \cite{klein} it was shown
that this relation implies that all constraints are satisfied.

\section{$Z_{\rm odd}$ simple current invariants} \label{sec-odd}

First we review some facts about simple
currents~\cite{simple}~\cite{intril}. An important quantity is the
(monodromy) charge of a field $i$ with respect to $J$
\begin{equation} \label{eq:charge}
	Q_J (i) = h_J + h_i - h_{J \times i} \;\; {\rm mod} \,1 \;\; ,
\label{eq:Qdefin}
\end{equation}
where the $h_i$ are the conformal weights of $i$. The order of a simple
current is the smallest integer $N$ for which $J^N=0$. The charge of a simple current with respect to itself is $Q_J (J) = r/N \;{\rm mod}\;1$ where $r$
is the monodromy parameter. In~\cite{simple} it is explained how we can
construct modular invariant partition functions with simple
currents\footnote{We multiply the modular invariants found in~\cite{simple} by the charge conjugation matrix $C_{ij}$. The result is of course a modular
invariant as well.}. The result for odd $N$ can be summarized as follows: the modular invariant is of pure automorphism type if and only if the monodromy parameter $r$ and the order $N$ do not have common factors.

All primaries in a theory with such a current are organized in orbits of
length $N$. The charges of these fields with respect to $J$ are multiples of  $r/N$, and every charge appears precisely once on a given orbit. We will
denote the charge-zero fields by $i_0$.

The following torus is modular invariant~\cite{simple}:
\begin{equation}\label{eq:torusp}
	T = \sum_{i_0} \sum_{n=1}^{N} \chi_{[J^ni_0]}
\bar{\chi}_{[J^ni_0^c]} \;\;\; .
\end{equation}
Note that the charge-zero fields are the transverse fields. The diagonal
part of this partition function is
\begin{equation} \label{eq:diaT}
	Z_{ii} = C_{i_0 i_0} \;\; ,
\end{equation}
where $C$ denotes the charge conjugation matrix. So a field $i$ appears
diagonally in the torus when the charge-zero field on the orbit it lies on is self-conjugate.

As an ansatz for the boundary and crosscap coefficients we
take~\footnote{This is a straightforward generalization of the results
of~\cite{class}. The classifying algebra is in our case just the charge-zero subalgebra of the fusion rules.}
\begin{equation} \label{eq:bp}
	B_{b\alpha} = \sqrt{N} \frac{S_{b\alpha}}{\sqrt{S_{b0}}} \;\;\; , \;\;\;
	\Gamma_b = \sqrt{N} \frac{P_{b0}}{\sqrt{S_{b0}}} \;\;\; ,
\end{equation}
where $b$ labels the transverse fields and the index $\alpha$ the boundary
conditions, which are in one-to-one correspondence with the orbits. Note that the boundary coefficient is independent of a representative, since $b$ is
chargeless and so $S_{b,J^t\alpha} = S_{b\alpha}$ for any $t$. It is
straightforward to compute the other coefficients
\begin{equation}
 	K_i  =  \sum_{n=1}^N Y_{J^ni,00} \label{eq:Kp} \;\;\; , \;\;\;
	A_{i\alpha\beta}  = \sum_{n=1}^N N_{J^ni,\alpha\beta} \;\;\; , \;\;\;
	M_{i\alpha} = \sum_{n=1}^N Y_{J^n\alpha,0i} \;\;\; .
\end{equation}
Note that in all three expressions only one term contributes. For instance
the Klein bottle coefficient can be written as $K_i = Y_{i_000}$, so it
equals the Frobenius-Schur indicator of the orbit to which $i$ belongs.
We will postulate that
nonzero Frobenius-Schur indicators are conserved in
fusion. Possible violations of this postulate and possible consequences
will be discussed in section  \ref{sec-cc}. If the postulate holds
the Klein bottle constraint~(\ref{eq:epsilon}) is satisfied.

With equation~(\ref{eq:bound}) it is straightforward to check positivity
and integrality of the open sector, equation~(\ref{eq:posintM}). Furthermore it may be shown that the annulus coefficient respects the completeness
conditions. This is strong evidence for the correctness of our
ansatz~(\ref{eq:bp}).

\subsection{Non-standard Klein bottles}

In~\cite{klein} we developed a method for constructing non-trivial Klein
bottles with the use of simple currents in the C-diagonal case. Simple
current Klein bottles can be constructed for $Z_{\rm odd}$ simple current
invariants as well. Let $L$ be an order $M$ simple current with arbitrary
spin. Then
\begin{equation}
	K^{[L]}_{i} = e^{2\pi i Q_L(i_0)} Y_{i_0 00} = Y_{i_0LL^c}
\end{equation}
is a consistent Klein bottle. Reality of this Klein bottle follows from the observation that $Q_L(i_0)$ is (half-)integer when $i_0$ is real. When $i_0$ is complex, a possible imaginary phase factor is killed by the
Frobenius-Schur indicator. The Klein bottle constraint~(\ref{eq:epsilon})
follows from the conservation of $Q_L$ in fusion. The other coefficients are
\begin{eqnarray} \label{eq:cp}
	&&B^{[L]}_{b\alpha} = \sqrt{N} \frac{S_{b\alpha}}{\sqrt{S_{bL}}} \;\; , \;\; \Gamma^{[L]}_b = \sqrt{N} \frac{P_{bL}}{\sqrt{S_{bL}}} \;\; , \;\; \\
 	&&A^{[L]}_{i\alpha\beta}  = \sum_{n=1}^N N_{[J^nL^ci],\alpha,\beta}
\;\; , \;\; 
	M^{[L]}_{i\alpha} = \sum_{n=1}^N Y_{[J^nL^c\alpha],L,i} \;\;\; .
\end{eqnarray}
It is straightforward to show that the various consistency conditions are
satisfied.

\section{$Z_2$ simple current invariants} \label{sec-two}

In this section we discuss the open descendants of modular invariants built from order $2$ simple currents. Only when the spin of the current is
half-integer is the resulting invariant a pure automorphism~\cite{simple}. There are three kinds of orbits: length $2$ orbits that contain charge-zero fields $i_0$, length $2$ orbits that contain charge-one-half fields $i_1$ and length $1$ orbits (fixed points) that contain charge-one-half fields $f$.
Let $J$ denote such a half-integer spin, order $2$ simple current. The
following torus is modular invariant~\cite{simple}:
\begin{equation}\label{eq:torus2}
	T = \sum_{i_0} \chi_{i_0} \bar{\chi}_{i_0^c} + \sum_{i_1} \chi_{i_1} \bar{\chi}_{(Ji_1)^c} + \sum_{f} \chi_f \bar{\chi}_{f^c} \;\; .
\end{equation}

The authors of~\cite{class} were able to find the boundary coefficients:
\begin{eqnarray}
	B_{b\alpha} =  \sqrt{2} \left ( \frac{S_{b\alpha}}{\sqrt{S_{b0}}} \right )	& & B_{b,f^{\pm}}  =  \frac{1}{\sqrt{2}}  \left (
\frac{S_{bf}}{\sqrt{S_{b0}}} \right )  \;\;\;\;\; {\rm for} \;\;\;\; Q(b)=0 \;\; ,\label{eq:B1} \\
	B_{b\alpha}  = 0 \;\;  \;\; & &
	 B_{b,f^{\pm}}  =  \pm \frac{1}{\sqrt{2}}  \left (
\frac{\breve{S}_{bf}}{\sqrt{S_{b0}}}\right )  \;\;\;\;\; {\rm for} \;\;\;\; Jb=b \;\; .\label{eq:B4}
\end{eqnarray}
where, in the case of WZW-models, 
$\breve{S}$ is the $S$-matrix of the fixed point conformal field
theory (FCFT)~\cite{simple}, or, more generally, the
orbit Lie-algebra \cite{FSSA}. In section \ref{breveS} we will
comment on its definition
in general conformal field theories.  
Note that there is one boundary coefficient
$\alpha$ for every length $2$ orbit and two boundary coefficients $f^{\pm}$ for every fixed point. As an ansatz for the crosscap coefficient we take
\begin{equation} \label{eq:Gamma}
	\Gamma_b =  \frac{1}{\sqrt{2}} \frac{[P_{b0} + \epsilon_{J,m}
P_{bJ}]}{\sqrt{S_{b0}}} \;\;\; ,
\end{equation}
where $\epsilon_{J,m} = e^{\pi i [h_J + \frac{m+2}{2}]} = \pm 1$, whose
appearance will become clear later. The odd integer $m$ is defined in
section \ref{breveS}. The Klein bottle becomes
\begin{equation} \label{eq:KB2}
	K_i = \frac{1}{2} [Y_{i00} + Y_{iJJ} + 2 \epsilon_{J,m} Y_{i0J}] \;\;\; .
\end{equation} 	
Using relations from the appendix of~\cite{klein} one can show that
for chargeless fields the Klein bottle is $Y_{i00}$ and for charged fields (including fixed points) it is $\epsilon_{J,m} Y_{i0J}$.
There are cases (although rare) where~(\ref{eq:KB2}) violates the Klein
bottle constraint~(\ref{eq:epsilon}). In section ~\ref{sec-cc} we will
comment on this issue and propose a slightly weaker version of the Klein
bottle condition that is satisfied by all models.
The annulus~\cite{class} and M\"obius coefficients become
\begin{eqnarray}
	& & A_{i\alpha\beta}  =  N_{i\alpha\beta}  + N_{Ji,\alpha\beta}
\;\;\;\; , \;\;\;\;
	A_{if^{\pm}g^{\pm}}  =  \frac{1}{2}[N_{ifg}  +  \breve{N}_{ifg}]
\;\;\;\; ,\\
	& & A_{i\alpha f^{\pm}}   =  N_{i\alpha f}\;\;\;\; , \;\;\;\;
	A_{if^{\pm}g^{\mp}}  =  \frac{1}{2} [N_{ifg}  -  \breve{N}_{ifg}]
\;\;\;\; , \\
	& & M_{i\alpha}  =  Y_{\alpha 0i} + Y_{\alpha Ji} \;\;\;\; , \;\;\;\;
	M_{if^{\pm}}  =   \frac{1}{2} [Y_{f0i}  + \epsilon_{J,m} Y_{fJi}]
\;\;\;\; ,
\end{eqnarray}
where $\breve{N}_{ifg}$ is defined in equation~(\ref{eq:cupN}). The annulus coefficients satisfy~\cite{class} the completeness conditions
of~\cite{completeness}. The positivity and integrality condition for boundary coefficients that are not fixed points follows from
equation~(\ref{eq:bound}). For boundary coefficients that are fixed points, we have to prove
\begin{eqnarray}
	\frac{1}{2} |Y_{f0i}  + \epsilon_{J,m} Y_{fJi}| & \leq &
\frac{1}{2}[N_{iff}  +  \breve{N}_{iff}]  \;\;\; ,\\
	\frac{1}{2} [Y_{f0i}  + \epsilon_{J,m} Y_{fJi}] & = &
\frac{1}{2}[N_{iff}  +  \breve{N}_{iff}]  \;\;\;\; {\rm mod} \;2 \;\;\; .
\end{eqnarray}
Due to the appearance of the $\breve{N}$, equation~(\ref{eq:bound}) is not
applicable in this case. However, note that for $i=(i_1,f)$ the above relations are easily satisfied, since both sides vanish by charge conservation.  In
appendix A we prove the above positivity and integrality condition for
$i=i_0$. Note that without the sign $\epsilon_{J,m}$ in
equation~(\ref{eq:Gamma}), the positivity and integrality condition will be violated for theories with $\epsilon_{J,m}=-1$.

\subsection{A non-standard Klein bottle}

In this section we include a non-standard Klein bottle to our solution. Let $L$ denote an arbitrary simple current. Consider
\begin{equation} \label{eq:KBL}
	K_i^{[L]} = e^{2 \pi i Q_L (i)} K_i \;\;\; .
\end{equation}
Note that not every $L$ is allowed. We have to require that $2 Q_L(i) = 0
\;{\rm mod} \; 1$ when $K_i \neq 0$, otherwise the Klein bottle coefficient is not real. Recall~\cite{simple} that the charge of a field with respect to {\rm any} simple current is (half-)integer when the field is self-conjugate. So the Klein bottle coefficient of a charge-zero field $i_0$ and a fixed
point $f$ is a sign for any $L$, since these fields only appear in the direct Klein bottle when they are self-conjugate. From the torus~(\ref{eq:torus2}), we see that charge-one-half fields $i_1$ propagate in the direct Klein bottle when they satisfy $i_1=Ji_1^c$, which implies, by charge conservation, $Q_L(J) = 2 Q_L(i_1)$. We conclude that
equation~(\ref{eq:KBL}) is a consistent Klein bottle if $Q_L(J) = 0 \;{\rm
mod} \; 1$. Note that there is at least one $L$ that satisfies this
requirement, namely $L=J$. Let us focus on this case. Then
equation~(\ref{eq:KBL}) becomes
\begin{equation}
	K_i^{[J]} = \frac{1}{2} [Y_{i00} + Y_{iJJ} - 2\epsilon_{J,m}
Y_{i0J}]\;\;\; .	
\end{equation}
The crosscap coefficient is
\begin{equation} \label{eq:GammaL}
	\Gamma_b^{[J]} =  \frac{1}{\sqrt{2}} \frac{[P_{b0} - \epsilon_{J,m}  P_{bJ}]}{\sqrt{S_{b0}}} \;\;\; .
\end{equation}
As an ansatz for the boundary coefficients, we take equations~(\ref{eq:B1}) and~(\ref{eq:B4}) and replace $\sqrt{S_{b0}}$ by $\sqrt{S_{bJ}}$ as
in~\cite{klein}. The annulus and M\"obius coefficients that change relative to the standard Klein bottle projection are
\begin{eqnarray}
	&&A_{if^{\pm}g^{\pm}}^{[J]}  =  \frac{1}{2}[N_{ifg} -
\breve{N}_{ifg}] \;\;\; , \;\;\;
	A_{if^{\pm}g^{\mp}}^{[J]}  =  \frac{1}{2} [N_{ifg} +
\breve{N}_{ifg}] \;\;\; , \;\;\;\\
	&&M_{if^{\pm}}^{[J]}  =   \frac{1}{2} [Y_{f0i} - \epsilon_{J,m}
Y_{fJi}] \;\;\;\; .
\end{eqnarray}
Miraculously, the positivity and integrality condition in the open sector
for fixed point boundary indices now becomes
\begin{eqnarray}
	\frac{1}{2} |Y_{f0i}  - \epsilon_{J,m} Y_{fJi}| & \leq &
\frac{1}{2}[N_{iff}  -  \breve{N}_{iff}]  \;\;\; ,\\
	\frac{1}{2} [Y_{f0i}  - \epsilon_{J,m} Y_{fJi}] & = &
\frac{1}{2}[N_{iff}  -  \breve{N}_{iff}]  \;\;\;\; {\rm mod} \;2 \;\;\; ,
\end{eqnarray}
which we prove in appendix A.

\subsection{The matrix $\breve{S}$}\label{breveS}

In WZW-models the matrix $\breve{S}$ is the modular transformation
matrix of the orbit Lie algebra, and it differs by a known phase
from the fixed point resolution matrix $S^J$. 
The fixed point resolution matrices are explicitly known for
WZW-models \cite{FSS} and extended WZW models \cite{WZWE}.
The generalization of $\breve{S}$ beyond WZW-models is not
straightforward, since orbit Lie algebras are related to foldings
of Dynking diagrams, a concept that has no obvious CFT
generalization. One could try to define $\breve{S}$ as the
transformation matrix of the twining characters (as defined in \cite{FSSA}),
but their definition does not straightforwardly
generalize either.
Even if such a generalization is possible,
we do not know
if the twining characters
are always well-behaved under modular transformations,  nor the
relation between   $\breve{S}$ and $S^J$. We will postulate here that
$\breve{S}$ exists and that it is related to $S^J$ in a similar way
as in WZW-models. Thus we take as the definition of $\breve{S}$
\begin{equation}\label{eq:phaserel}
\breve{S} = e^{ 6 \pi i \frac{m}{24}} S^J\ . 
\end{equation}
For (half)-integer spin currents of WZW-models the number $m$
(which is defined modulo 24 and is related to a phase-shift in the matrix $T$)
is known, and it is
an even (odd) integer for  integer (half-integer)
spin currents.
This is what we will assume in
general.  Note that for half-integer spin currents there
are twelve possible values for $m$, resulting in four possible
phases in the definition of $\breve{S}$. Without further information
$\breve{S}$ is then known up to a factor $\pm 1$ or $\pm i$. In our
case a sign change is irrelevant, since this only determines the
choice between $f^+$ and $f^-$. A change by a factor $i$ changes the
sign of $\epsilon_{J,m}$  and $\breve{N}$. These sign changes can be
flipped by choosing the opposite Klein bottle projection, as explained
above, so that finally there is no genuine ambiguity left.
This definition of $\breve{S}$ appears to be adequate in all cases
with the possible exception of the Klein bottle constraint violations
mentioned in the next subsection.

\subsection{The Klein bottle constraint} \label{sec-cc}

In this section we discuss the Klein bottle constraint for open descendants of order $2$ automorphism invariants. This condition reads
\begin{equation}
	N_{ij}^{~~k} K_i K_j K_k \geq 0 \;\;\; ,
\end{equation}
where the Klein bottle for the various fields is
\begin{equation} \label{eq:Ktwo}
	K_{i_0} = Y_{i_000} \;\;\; , \;\;\; K_{i_1} = \epsilon_{m,J}
Y_{i_10J} \;\;\; , \;\;\; K_{f} = \epsilon_{m,J} Y_{f0J}=\eta_f Y_{f00}
\;\;\; ,
\end{equation}
where in the last step (\ref{eq:exp}) was used.

Since the charges with respect to any simple current are conserved (mod $1$) in fusion, there are four sectors where the fusion coefficients are
non-zero. We will discuss the Klein bottle constraint in these sectors one by one. \begin{itemize}
	\item {\bf The coupling between three charge-zero fields.}

 Since the Klein bottle in this sector is just the Frobenius-Schur indicator (in the
original theory), the Klein bottle constraint is satisfied trivially.
	\item {\bf The coupling between a charge-zero field and two fixed
points.} 

{}From equation~(\ref{eq:Ktwo}) we find that
the Klein bottle coefficient for a fixed point is
equal to its Frobenius-Schur indicator times a sign, $\eta_f$.
If $\eta_f$ does not depend on $f$ the Klein bottle constraint
is satisfied (assuming it holds in the C-diagonal case).
In case when two self-conjugate fixed points $f$ and $g$ with $\eta_f \neq
\eta_g$ have a nonzero coupling with a charge-zero field $i_0$, the Klein
bottle constraint is violated! We will comment on this below.
\end{itemize}

\noindent For the remaining two cases we make use of the extended tensor
product method of appendix A. We tensor the theory under
consideration with a second one, also with a half-integer spin
current, and extend the chiral algebra by the product of the 
half-integer spin currents of the two theories.
It is convenient to tensor with a series of theories , namely
$SO(2N+1)$ level 1. In these theories
$Y_{\bar f\bar 0\bar J}=(-1)^N  Y_{\bar f\bar 0\bar 0}
 \equiv \epsilon_N$, which is a sign. Then we derive from
equations  (\ref{eq:YeZero}),  		
(\ref{eq:YeOne}) and (\ref{eq:YeFixed})\begin{eqnarray}
	Y_{i_e0_e0_e} & = & Y_{i_000}\label{eq:YeIsingZero} \\ 		
	Y_{j_e0_e0_e} & = & \epsilon_N [Y_{j_100} + (-1)^N \epsilon_J
Y_{j_10J}] \label{eq:YeIsingOne} \\
	Y_{f_{e,\mu}0_e0_e} & = & {1\over 2}\epsilon_N[Y_{f00}
+ (-1)^N \epsilon_J Y_{f0J}]
=
{1\over 2}\epsilon_N K_f [\eta_f + \epsilon_J(-1)^N\epsilon_{J,m}]
\label{eq:YeIsingFixed}
\end{eqnarray}
where $\epsilon_J = e^{\pi i [h_J + \frac{1}{2}]}$. Furthermore we have	
\begin{equation} \label{eq:NeIsing}
	N_{i_ej_e}^{\;\;\;k_e} = N_{i_0j_1}^{\;\;\;k_1} +
N_{Ji_0,j_1}^{\;\;\;k_1} \;\;\; ,\;\;\;
	 N_{i_ej_e}^{\;\;\;f_{e,\mu}}  =  N_{i_0j_1}^{\;\;\;f}
\end{equation}
Now we use the Klein bottle constraint in the extended tensor product
theory to deal with the last two cases.
\begin{itemize}
	\item {\bf The coupling between a charge-zero field and two
charge-one-half fields.}

We wish to prove
\begin{equation}
	N_{i_0j_1}^{\;\;\;k_1}K_{i_0} K_{j_1} K_{k_1} \geq 0 \;\;\; .
\end{equation}

{}From the torus~(\ref{eq:torus2}), we know that a charge-one-half field $j_1$ propagates in the direct Klein bottle when $j_1^c = Jj_1$. So these fields must be complex, {\it i.e.},  have a vanishing Frobenius-Schur indicator $Y_{j_100}$. From equations~(\ref{eq:YeIsingOne}) and~(\ref{eq:Ktwo}), we see that the Klein bottle coefficient for these fields becomes
$K_{j_1}=(-1)^N \epsilon_{J,m} \epsilon_N \epsilon_J Y_{j_e0_e0_e}$.
However, the signs are irrelevant, since they do not depend on $j_1$ and
cancel between the two charge-one-half fields. Then the Klein
bottle constraint follows from the (proposed) conservation of Frobenius-Schur indicators in the extended tensor theory
\begin{equation}
	N_{i_ej_e}^{\;\;\;k_e}Y_{i_e0_e0_e}Y_{j_e0_e0_e}Y_{k_e0_e0_e} \geq 0
\end{equation}
and the fact that $N_{i_ej_e}^{\;\;\;k_e} \geq N_{i_0j_1}^{\;\;\;k_1}$,
as a consequence of equation~(\ref{eq:NeIsing}).
	\item {\bf The coupling between a charge-zero field, a charge-one-half field and a fixed point.}

We can always choose $N$ in (\ref{eq:YeIsingFixed}) such that the
two terms do not cancel. For
that choice of $N$ we have then
$K_{f}=(-1)^N \epsilon_{J,m} \epsilon_N \epsilon_J Y_{f_{e,\mu}0_e0_e}$,
exactly the relation we found in the previous case. Then the fixed point
behaves like an ordinary charge-one-half field, and the rest of the argument
is completely analogous to the previous case.

\end{itemize}

\subsection{Violations of the Klein bottle constraint}

The violation of the Klein bottle constraint 
noted above may be resolved in several
ways. First of all it is possible that either such models must be
rejected, or that there is something wrong with the general formalism
developed here. In the examples where the violation occurs we always
have to use the conjectured relation (\ref{eq:phaserel})
 to compute $\breve{S}$.
This may cast some suspicion on this conjecture. However, there is
a more interesting possibility. 
The violation is due to different signs $\eta_f$ in the
relation $\breve{S}^2_{fg}=\eta_f C_{fg}$. This splits the set of
fixed points into two sets, one with $\eta_f=1$ and one with $\eta_f=-1$.
Conjugation closes on each set.
It is easy to show that  $\breve{S}_{fg}$ must have vanishing matrix
elements between fields of different sets. This then implies that
$\breve{N}_{ifg}=0$. This immediately implies 
 that $N_{ifg}$ must be even, since the average
of both coefficients must be an integer \cite{class}.
This may point towards
a possible solution of the puzzle. 

It is instructive to consider the postulate of section 3
regarding conservation of the Frobenius-Schur indicator in fusion. 
The FS-indicator is the generalization to CFT of the notion of
real (R, FS=1), pseudo-real (P, FS=$-1$) 
or complex representations (C, FS=0) \cite{bantay}. This
property is preserved in 
simple Lie-algebra tensor products: schematically $R \times R = R$,
$P \times P = R$ and $P \times R = P$, where on the left-hand
side we consider two irreducible representations and on the right-hand
side we have a direct sum of irreducible representations, which may
consist out of complex conjugate pairs. Strictly speaking the latter
already violate FS-conservation, but since their indicator is 0 such a
violation does not affect the Klein bottle constraint.  
By analogy with
FS-indicator of simple Lie-algebra representations, we expect the
FS-indicator to be preserved in fusion unless the fusion 
produces a pair of complex conjugate representations (which together
can form a real or pseudo-real representation) or an even number
of real or pseudo-real representations. The latter possibilities
are logically possible (for example for two pseudo-real representations
one can choose a real basis), but they never occur in WZW-fusion or
simple Lie-algebra tensor products because there always exists a
conserved simple current charge which is 0 on real representations,
$1/2$ on pseudo-real ones and anything on complex representations. The
conservation of this charge then enforces conservation of the FS-indicator
in the sense described above. However, in general CFT's we may expect
violations of the conservation rule, leading to a violation of the
Klein bottle constraint (\ref{eq:epsilon}) by negative, {\it even} integers. 
We do not know of any CFT where such a violation occurs for
the FS-indicators ({\it i.e.} for the Klein bottles of the C-diagonal 
models), but remarkably we find precisely such a violation for the
Klein bottles of certain off-diagonal models. 

This leads one to suspect that perhaps violations by even negative
integers are allowed. Consider for example
two fields with Klein bottle coefficients
$K_1=K_2=1$ and a third with
$K_3=-1$. The existence of a closed string coupling of the first two
fields to the third implies a coupling of two symmetrically projected
unoriented string fields to the symmetric  projection of the third one.
But if $K_3=-1$ the symmetric projection is an unphysical state, since
the Klein bottle produces an anti-symmetric projection.
 However, if the coupling has multiplicity two (or even) it is possible for
fields 1 and 2 to couple to a symmetric combination of two 
anti-symmetrically projected (hence physical) states 3. This  is 
precisely the reason why a violation of FS-conservation was argued
to be possible in principle, and is presumably how these Klein bottle
constraint violations should be interpreted as well. If this 
interpretation is correct there is no inconsistency, and the
formulation of the Klein bottle constraint must be weakened as
explained above.    

\section{Conclusions and outlook}

In this paper we have presented natural and general candidates
for crosscap coefficients belonging to theories with
a non-trivial automorphism invariants in the bulk. Although we
do not pretend to have proved that the resulting open string theories
are fully consistent, at least we have demonstrated that our
solution passes a number of non-trivial checks. In rare cases 
there may be a problem with the Klein bottle constraint, but we
have argued that the fact that the violation is always by even
integers may provide a way out of the problem, and that perhaps
a slightly weaker form of the constraint might be acceptable. 

The crosscap coefficients are an essential ingredient in 
the tadpole cancellation conditions, which can now be studied
for open descendants of 
a much larger class of  bulk theories. 

Nevertheless there are still several classes of automorphism 
invariants for which the crosscap coefficients remain unknown.
Obvious extensions of our results can be made to 
automorphisms generated by $Z_{2^n}$ simple currents and to non-cyclic
simple current groups. One may in fact hope for one general formula
covering all cases. We expect to address this question in the
near future. Another interesting direction is the study of
diagonal invariants ({\it i.e.} a charge conjugation invariant of
a C-diagonal theory) and exceptional invariants such as the ``$E_7$"
invariant of $SU(2)$ level 16.
Furthermore the entire formalism should be extended to the 
case of boundaries that break bulk symmetries.

\appendix

\section{Inequalities from tensoring} \label{sec-cons}

In this appendix we obtain some useful relations among the quantities
$N, \breve{N}$ and $Y$, which are needed for proving integrality,
positivity and the Klein bottle constraint.

The strategy is as follows. We will show that the Klein bottle coefficient
of section~\ref{sec-two} can be related to the Frobenius-Schur indicator of a different theory. The conservation of this indicator will translate into the Klein bottle constraint. Similarly, equation~(\ref{eq:bound}), which holds
in this theory, turns out to prove positivity and integrality in the open
sector.

More explicitly, we tensor the theory of interest with another
theory that contains a half-integer spin simple current. We then extend the chiral algebra of this tensor theory with the products of the half-integer
spin simple currents. We label the fields in the tensor product by $i_t =
(i,\bar{i})$ and
the fields in the extended tensor product by $i_e$.
In particular, $J_t=(J,\bar{J})$ is a simple current and when $f$ is a fixed point of $J$ and $\bar{f}$ a fixed point of $\bar{J}$, $f_t =(f,\bar{f})$ is a fixed point of $J_t$.

The matrices $S,T$ and $P$ and the coefficients $N_{ijk}$ and $Y_{ijk}$
of the tensor theory are related to those of its components
in a straightforward way. To deal with the extension we make use
of the results of \cite{FSS}\  to resolve the fixed points of $J_t$. This
then gives us the matrix $S$ of the extended theory in terms of the
matrices $S$ and $S^J$ (the fixed point resolution matrix) of the
components. We then compute $P$, $N$ and $Y$.

In the extended theory only fields $i_t =(i,\bar{i})$ that are chargeless
under $J_t$ exist, so $Q_J(i)=Q_{\bar{J}}(\bar{i})$. Furthermore, these
fields are grouped by the action of the simple current $J_t$.
For every fixed point in the tensor theory there are two fields in the
extension~\cite{simple}, which we label by $\mu, \nu = (1,2)$.

The matrices $S,T$ and $P$ of this extension are related to those of the
unextended theory, which is the tensor theory. For the matrix $S$ we have for instance
\begin{equation}
	S_{i_ej_e} = 2S_{i_tj_t} \;\;\; , \;\;\; S_{i_ef_{e,\mu}} =
S_{i_tf_t} \;\;\; .
\end{equation}
When the indices are two fixed points the result is
\begin{equation}
	S_{f_{e,\mu}g_{e,\u}} = S_{f_tg_t} E_{\mu\u} +
S^J_{f_tg_t} F_{\mu\u} \;\;\; ,
\end{equation}
where the matrices $E$ and $F$ are given by
\begin{equation}
	E = \left (	\begin{array}{cc}
			1 & 1 \\
			1 & 1
		\end{array} \right )
\;\;\;\; , \;\;\;\;
	F = \left (	\begin{array}{rr}
			1 & -1 \\
			-1 & 1
		\end{array} \right ) \;\;\; .
\end{equation}

The matrix $P$ in the extension, is given by
\begin{equation}
	P_{i_ej_e} = 2 \bar{P}_{i_tj_t} \;\;\; , \;\;\; P_{f_{e,\mu}j_e} =
\bar{P}_{f_tj_t} \;\;\; , \;\;\; P_{f_{e,\mu}g_{e,\u}} = 0 \;\;\; .
\end{equation}
where the ``orbit averaged $P$-matrix'' is ($x=i_t, f_t$)
\begin{equation}
	\bar{P}_{xj_t} = \frac{1}{2} \left [ P_{xj_t} +
\sqrt{\frac{T_{j_tj_t}}{T_{Jj_t,Jj_t}}} P_{xJj_t} \right ] \;\;\; .
\end{equation}

We can now relate the tensor $Y$ of the extended theory to the tensors $Y$
of the components of the tensor product. We will be particularly interested in the Frobenius-Schur indicators of $i_e = (i_0,\bar{0}) + (Ji_0,\bar{J})$, $j_e = (j_1,\bar{f}) + (Jj_1,\bar{f})$ and $f_e = (f,\bar{f})$. The
subscripts $0$ and $1$ were introduced in section~\ref{sec-two}, and refer to the charges of the fields. A straightforward calculation shows
\begin{eqnarray}
	Y_{i_e0_e0_e} & = & Y_{i_000} Y_{\bar{0}\bar{0}\bar{0}} +
\epsilon_{J,\bar{J}} Y_{i_00J} Y_{\bar{0}\bar{0}\bar{J}} =
Y_{i_000} \;\;\; ,
 \label{eq:YeZero}\\ 		
	Y_{j_e0_e0_e} & = & Y_{j_100} Y_{\bar{f}\bar{0}\bar{0}} +
\epsilon_{J,\bar{J}} Y_{j_10J} Y_{\bar{f}\bar{0}\bar{J}} \;\;\; ,
\label{eq:YeOne}\\
	Y_{f_{e,\mu}0_e0_e} & = & {1\over 2}[Y_{f00}
Y_{\bar{f}\bar{0}\bar{0}} + \epsilon_{J,\bar{J}} Y_{f0J}
Y_{\bar{f}\bar{0}\bar{J}}] \label{eq:YeFixed} \;\;\; ,
\end{eqnarray}
where $\epsilon_{J,\bar{J}} = e^{\pi i [h_J + h_{\bar{J}}]} = \pm 1$.
In the first line we used $|Y_{\bar{0}\bar{0}\bar{J}}|
\leq N^{\bar J}_{~\bar{0}\bar{0}} = 0$ (see equation~(\ref{eq:bound})).
The fusion coefficients in the extended theory can also be related to those of the components of the tensor theory. We have for example:
\begin{eqnarray}
	N_{i_ej_e}^{\;\;\;k_e} & = &
N_{i_0j_1}^{\;\;\;k_1}N_{\bar{0}\bar{f}}^{\;\;\;\bar{f}} +
N_{Ji_0j_1}^{\;\;\;k_1}N_{\bar{J}\bar{f}}^{\;\;\;\bar{f}} \;\;\; , \\
	 N_{i_ej_e}^{\;\;\;f_{e,\mu}} & = &
N_{i_0j_1}^{\;\;\;f}N_{\bar{0}\bar{f}}^{\;\;\;\bar{f}} \;\;\; ,
\end{eqnarray}
where $k_e = (k_1,\bar{f}) + (Jk_1,\bar{f})$.

One can also show that
\begin{equation} \label{eq:Y}
	Y_{f_{e,\mu}0_ei_e} = \frac{1}{2} [Y_{f0i_0}Y_{\bar{f}
\bar{0}\bar{0}} +  \epsilon_{J,\bar{J}} Y_{fJi_0}Y_{\bar{f} \bar{J}\bar{0}}]
\end{equation}
and 	
\begin{equation} \label{eq:N}
	N_{i_ef_{e,\mu}g_{e,\mu}} = \frac{1}{2}[
N_{i_0fg}N_{\bar{0}\bar{f}\bar{g}} + e^{-\frac{1}{2} \pi i (m+\bar{m})}
\breve{N}_{i_0fg}\breve{N}_{\bar{0}\bar{f}\bar{g}}] \;\;\; ,
\end{equation}
where
\begin{equation} \label{eq:cupN}
	\breve{N}_{ifg} = \sum_{h, Jh=h} \frac{S_{hi} \breve{S}_{hf}
\breve{S}_{hg}}{S_{h0}} \;\;\; , \;\;\; \breve{N}_{0fg} = \eta_f C_{fg}
\;\;\; ,
\end{equation}
where $\eta_f$ is a sign. The phase factor in (\ref{eq:N}) is due to the
fact that we have expressed the fixed point resolution matrices 
$S^J$ (which appear naturally in fixed point resolutions for integer
spin currents) in terms of $\breve{S}$ using (\ref{eq:phaserel}). 
In the tensor theory
$m_t$ is an even integer, which decomposes as $m_t = m + \bar{m}$, where $m$ and $\bar{m}$ are odd.

In WZW-models this sign is in fact independent of $f$, but in extended
WZW-models examples are known where it is not~\cite{WZWE}.\footnote{Note
that $\eta$ is defined here in terms of the square of $\breve{S}$, not the
square of $S^J$ as in \cite{FSS}. If (\ref{eq:phaserel}) holds,
 $\eta_f=\pm 1$.}
In the extended  tensor theory, equation~(\ref{eq:bound}) must hold. So for instance
\begin{equation}\label{eq:FSconsE}
	|Y_{f_{e,\mu}0_ei_e}| \leq N_{i_ef_{e,\mu}f_{e,\mu}} \;\;\; .	
\end{equation}
We now make a specific choice for the ``bar-theory'', namely the Ising model with fields $\bar{0}=0$, $\bar{f} = \sigma$ and $\bar{J} = \psi$.
Equation~(\ref{eq:FSconsE}) then becomes
\begin{equation} \label{eq:bert}
 	\frac{1}{2} |Y_{f0i_0} \pm \epsilon_{J,m}  Y_{fJi_0}| \leq
	\frac{1}{2} [N_{i_0ff} \pm \breve{N}_{i_0ff}] \;\;\; ,
\end{equation}
where $\epsilon_{J,m}$ is defined in the main text. Note that the above  
equations are the positivity conditions in the open sector where the boundary 
labels are fixed points for the two Klein bottle projections of
section~\ref{sec-two}. Integrality can be proved in a similar way. It is not hard to convince oneself that the above bound is independent of the chosen
``bar-theory''. Another consequence of this inequality can be seen by taking $i=0$. With the use of equation~(\ref{eq:cupN}) we get
\begin{equation}
 	\frac{1}{2} |Y_{f00} \pm \epsilon_{J,m}  Y_{fJ0}| \leq
	\frac{1}{2} C_{ff} \pm \eta_f C_{ff} \;\;\; .
\end{equation}
Note that this implies
\begin{equation} \label{eq:exp}
	Y_{f00} = \eta_f \epsilon_{m,J} Y_{f0J} \;\;\; .
\end{equation}

\vspace{10mm}

\begin{center}
{\bf Acknowledgements}
\end{center}

A.S. would like to thank Christoph Schweigert and J\"urgen Fuchs for discussions. L.H. would like to thank the 
``Samenwerkingsverband Mathematische Fysica" for financial support and
N.S. would like to thank Funda\c c\~ao para a Ci\^encia e Tecnologia for
financial support under the reference BD/13770/97.

\end{document}